\documentclass{proceedingsM}

\usepackage[utf8]{inputenc}

\usepackage{amssymb,amsmath,amsthm}
\usepackage{color}
\usepackage{epstopdf}

\usepackage[title]{appendix}

\usepackage{float} 

\usepackage[authoryear]{natbib}
\usepackage{wrapfig}

\DeclareMathOperator{\sech}{sech}

\title{Electromechanical coupling of waves in nerve fibres}
\author{J\"uri Engelbrecht, Tanel Peets and Kert Tamm}
\address{Laboratory of Solid Mechanics, Department of Cybernetics, School of Science, Tallinn University of Technology \\Akadeemia tee 21, Tallinn 12618, Estonia}
\email{je@ioc.ee, tanelp@ioc.ee, kert@ioc.ee}
\abstract{The propagation of an action potential (AP) in a nerve fibre is accompanied by mechanical and thermal effects. In this paper an attempt is made to build up a mathematical model which couples the AP with a possible pressure wave (PW) in the axoplasm and waves in the nerve fibre wall (longitudinal - LW and transverse - TW)  made of a lipid bilayer (biomembrane). A system of differential equations includes the governing equations of single waves with coupling forces between them. The single equations are kept as simple as possible in order to carry out the proof of concept. An assumption based on earlier studies is made that the coupling forces depend on changes (the gradient, time derivative) of the voltage. In addition it is assumed that the transverse displacement of the biomembrane can be calculated from the gradient of the LW in the biomembrane. The computational simulation is focused to determining the influence of possible coupling forces on the emergence of mechanical waves from the AP. As a result, an ensemble of waves (AP, PW, LW, TW) emerges. The further experiments should verify assumptions about coupling forces. In the Appendix, the numerical scheme used for simulations, is presented.}
\date{February 20, 2018}

\begin{document}
\maketitle

\section{Introduction}

The increasing understanding on complexity has influenced many fields of research. The role of coupling, interaction, self-organization, hierarchies, etc. in complex systems has lead to better understanding of natural and man-made systems and processes. Without any doubt this concerns also biosystems including biophysics, biochemistry, medical physics, genetics, etc. One of the fascinating problems of biophysics is the propagation of signals in nerve fibres and networks. The basic element of this process is the propagation of an electrical signal in a single axon. An important step in understanding this process was related to the studies of \citet{Hodgkin1945}, who derived a mathematical model of an axon potential (AP) based on the ionic hypothesis. The celebrated Hodgkin-Huxley (HH) model describes an electrical signal in a fibre that has a typical asymmetric shape and is strongly supported by ion currents through the fibre wall. Later several simplified models were proposed for the description of this process like the widely used FitzHugh-Nagumo (FHN) model \citep{Nagumo1962}. Instead of sodium and potassium ion currents governed by specific kinetic equations in the HH model, in the FHN model just one unspecified ion current is used which is able to reproduce the main properties of an AP. Paying full credit to  the HH model, one has to admit that the existence of more than 20 parameters in the model makes its practical usage difficult.

However, the process is more complicated than a single wave. Hodgkin himself stated that ``in thinking about the physical basis of the action potential perhaps the most important thing to do at the present moment is to consider whether there are any unexplained observations which have been neglected in an attempt to make the experiments fit into the a tidy pattern'' \citep{Hodgkin1964a}. Indeed, the structure of a nerve fibre is complicated \citep{Debanne2011}. Even in the first approximation it must be described as a tube surrounded by extracellular fluid with a wall made of a biomembrane and filled with the intracellular fluid – the axoplasm. The AP is an electrical signal in the axoplasm. The biomembrane is made of the lipid bilayer consisting of amphiphilic molecules with hydrophobic tails directed toward the membrane centre. This bilayer has also embedded proteins which are responsible for forming the ion transport channels (gates) through the membrane. Consequently, in order to get a full description of a process in a nerve fibre, in addition to the propagation of an AP, the accompanying processes in the surrounding biomembrane and in the axoplasm must be understood. 

In this paper a mathematical model describing coupled processes in a nerve fibre is presented. The resulting ensemble of waves is a clear sign of complexity of the process when the single constituents form a whole. In Section 2 the physical description of the general signal propagation in nerve fibres is revisited in order to formulate the background. The next Section 3 is devoted to the analysis of coupling mechanisms and known models. A novel model on the basis of governing equations and coupling forces is presented in Section 4. Attention is paid to assumptions and to the differences and similarities with known models. In Section 5 the results of the numerical simulation are presented. The final Section 6 involves conclusions and ideas for further theoretical and experimental studies. In the Appendix, the numerical scheme used for simualations, is described. 

\section{Physical description of signal propagation}

\paragraph{Early descriptions} Electrophysiology of nerves is strongly influenced by explanations given by Nernst in the beginning of the 20th century (see overview by \citet{Faraci2013}) who has described the movement of ions in nerve fibres. Even before Hodgkin and Huxley studies, \citet{Wilke1912}, \citet{Cole1939} have noted the complicated nature of signals in nerve fibres. \citet{Kaufmann1989} has analyzed the possible coupling effects: “electrical action potentials are inseparable from force, displacement, temperature, entropy and other membrane variables”. Indeed, nowadays several experiments have proved the existence of phenomena accompanying the propagation of an AP. That all has been summed up by the following statement: “... to frame a theory that incorporates all observed phenomena in one coherent and predictive theory of nerve signal propagation” \citep{Andersen2009}.

\paragraph{Experimental evidence} The early experiments of \citet{Hill1936} and \citet{Hodgkin1945,Hodgkin1964a} have demonstrated the formation of an AP in dependence of ion currents. In addition, the heat production associated with an AP was measured \citep{Abbott1958}. All this basic knowledge was summarized in \citep{Hodgkin1964a,Katz1966}. Later a lot of attention was focused also to mechanical effects accompanying the AP.  The swelling effects of a nerve fibre have been demonstrated \citep{Iwasa1980,Tasaki1989} and the pressure waves in axoplasm analyzed \citep{Terakawa1985}. It means that the main components of accompanying effects -- mechanical waves in the fibre wall and in the intracellular axoplasm generated due to the coupling with electrical signals have been experimentally measured. The observable transverse displacement of the biomembrane has been measured being about 1-2 nm. The overviews of these studies have summarized the findings \citep{Tasaki1988,Kaufmann1989}. Recent experimental studies have given more information about the dependence of those effects on physiological parameters \citep{Gonzalez-Perez2016}. The similar coupling of electrical signals and mechanical pulses have been measured also in excitable plant cells \citep{Fillafer2017}. 

\paragraph{Present understanding} Axon physiology is nowadays very well documented \citep{Clay2005,Debanne2011}. The axon walls are biomembranes which are specific lipid bilayers with many embedded cellular and molecular components which regulate the forces and transmission between the membrane and the ion channels \citep{Mueller2014}. Biomembranes are important not only for nerve fibres but also because biomembranes are structures characteristic to all living cells. These layers between living cells and the surrounding environment can be treated as deformable structures and they are able to carry mechanical waves. Consequently, the methods from the theory of continua (microstructured media) can be applied for deriving the governing equations of deformation. The mechanical energy of a biomembrane has been proposed as a quadratic function called after Helfrich and it describes the lipid bilayer as a homogeneous elastic body, usually two-dimensional \citep{Helfrich1973}. This approach is nowadays modified \citep{Lomholt2006,Deseri2008} and is able also to describe inhomogeneities of the biomembranes \citep{Bitbol2011}. These inhomogeneities are related to channels for ion transport \citep{Mueller2014}. An important proposal to account for physical nonlinearities gives a possibility to model localized waves in a biomembrane \citep{Heimburg2005}. The corresponding mathematical model is a Boussinesq-type equation \citep{Heimburg2005,Engelbrecht2015}. The electrical signal in an axon propagates in the intracellular axoplasm which is actually a gel consisting 87\% of water held together with cytoskeleton \citep{Gilbert1975}. The axoplasm is able to carry also the pressure waves \citep{Terakawa1985,Rvachev2010}.

All mechanisms, structural properties and models described briefly above reflect the specific features of signal propagation in nerve fibres. It is a challenge to build up a coupled model of all observable effects into one, much in terms of A.Toffler – “...we often forget to put the pieces back together again” \citep{Toffler1984}. Here it means that an AP should be coupled to mechanical deformation in the biomembrane and the pressure changes in the axoplasm.

\section{Modelling of mechanisms of coupling}

\paragraph{Open questions and difficulties} Although there is a general agreement that all the dynamical changes in nerve fibres during the propagation of an AP are coupled, the mechanisms of coupling are not satisfactorily described. This concerns the coupling between three main processes: AP, waves in biomembrane (LW,TW), and pressure waves (PW) in axoplasm. Taken as single processes, the physical and mathematical descriptions of them are well known.  As far as the AP is supported by ion currents, the transport through ion channels is also well studied \citep{Heimburg2010,Howells2012,Mueller2014}. It must be stressed that the ion channels may be influenced not only by electrical factors (voltage-gated) like in the HH model but may be also mechanically sensitive \citep{Mueller2014}. The opening of an ion channel means also the deformation of the lipid bilayer and once this is a time-dependent event, it produces a mechanical wave in the bilayer. This process has a localized character and the crucial problem is to understand the electromechanical transduction mechanisms – the transduction of electrical energy to mechanical (AP to waves in biomembrane) and vice versa. However, it is not clear yet whether electrostriction or piezoelectricity is the main mechanism of the transduction \citep{Gross1983}. Electrostriction is related to electric field-induced deformation in dielectrics and the produced stress is proportional to the square of the imposed electric field \citep{Gross1983} and seems to be a better candidate for coupling because based on known data, piezoelectricity leads to unrealistic values of deformation. Here, as suggested by \citet{Gross1983}, studies on molecular mechanisms in biomembranes should clarify the effects. Later, based on experiments, it has been stated that the mechanical changes in the biomembrane are proportional to the voltage changes \citep{Gonzalez-Perez2016}. However, it has been also argued that the mechanical effects in the biomembrane accompanying the AP could be caused by water movement associated by sodium influx through ion channels \citep{Kim2007}.  It is clear that the extracellular and intracellular molecular structure of a fibre has great impact on processes. The heat production accompanying the AP has noted in several studies \citep{Howarth1968,Heimburg2007,Gonzalez-Perez2016} and this process is seemingly responsible for phase changes in the lipid bilayer. That is why the absence of thermodynamical considerations in the HH model has been criticized compared to the adiabatic theories \citep{Heimburg2005,Gonzalez-Perez2016}.

An important question is related to the velocities of all the processes. Experimental studies have demonstrated that the estimations for velocities of single waves can be significantly different. The velocities of a nerve pulses depend on the diameter of fibres (but also on temperature, ion concentration, myelin thickness, etc.) and for human nerves can be in the interval starting from ca 2~m/s for nerves with a small diameter up to 100~m/s in bigger nerves \citep{Debanne2011}. The classical results of HH model give the estimation of about 20~m/s for the non-myelinated squid axon \citep{Hodgkin1964a}. In myelinated nerves, the velocities are larger \citep{Heimburg2005}. The estimations for localized mechanical waves in biomembranes indicate the values of velocity about 170~m/s \citep{Heimburg2005}. The velocities in excitable plant cells of electrical and mechanical waves both, however, have shown synchronization \citep{Fillafer2017} but are considerably slower (less than 10~m/s).  

The pressure waves in axoplasm can be analyzed like pulses in flexible tubes and then the velocities are dependent on viscosity of the intracellular fluid and the diameter but also on temperature \citep{Rvachev2010}. These theoretical estimations cover a wide interval from small velocities around several m/s up to velocities around 90~m/s. For modelling the pressure waves it is possible to use either the Navier-Stokes model or the direct analogy to the waves in tubes \citep{Engelbrecht2018}. One possible starting point is the two-dimensional (2D) model of pressure waves in a elastic cylindrical tube \citep{Lin1956}

	\begin{align}
	\label{LinMorganPressureWaves}
		&\bar{p}_{tt}=c_f^2(\bar{p}_{xx}+\bar{p}_{rr}+\bar{p}_r/r),\\
		&\rho u_{tt}+\bar{p}_x=0,\\
		&\rho w_{tt}+\bar{p}_r=0,
	\end{align}
where $\bar{p}$ is the pressure, $x$ and $r$ are the longitudinal and radial coordinates respectively; $u$, $w$ are the longitudinal and radial displacements respectively and $c_f$ is the velocity of sound in the fluid. Here and further, the independent variables used as indices, denote differentiation. As far as the diameter of the axon is very small, it is assumed that the pressure is constant across its cross-section ($\bar{p}_r=0$). In this case Eq.~\eqref{LinMorganPressureWaves} reduces to 
\begin{equation}
	\label{LinMorganWaveEQ}
	\bar{p}_{tt}=c_f^2\bar{p}_{xx}, 
\end{equation} %\quad {\color{blue}\rho u_{tt}+\bar{p}_x=0,}
which is the classical wave equation. Although Eq.~\eqref{LinMorganWaveEQ} does not include viscosity it is straightforward to take into account if needed by adding viscous damping term. The effects of nonlinearity are also not included because the amplitude of a pressure wave is small \citep{Terakawa1985}.

To sum up, in experiments more emphasis is directed towards the voltage of APs (amplitudes) rather than to velocities. It is clear that in a coupled model all the velocities should be synchronized. One must also stress that changes in nerve fibre properties are strongly influenced by anaesthetics \citep{Heimburg2007} that could influence the amplitudes and velocities of waves by changing the properties of lipid membranes, i.e., the ion transport.

\paragraph{Modelling of coupling} Without any doubt, there is a considerable interest to build up theories where at least basic effects are described within one model \emph{resp} theory. In many cases the models are formulated at the physical level determining the possible linkage of effects together with possible coupling factors. However, the approach where such a description is supported by mathematical models like the basic models describing single effects, seem to be more perspective. 

\citet{Hady2015} have proposed a model for coupling the electrical and mechanical
signals which is based on the assumption that the potential energy is mostly stored in the surrounding biomembrane and the kinetic energy in the axoplasmic fluid resulting in mechanical surface waves in the biomembrane. The AP is described by using the HH model and the force exerted on the biomembrane is taken proportional to the square of the voltage. The process in the axoplasmic fluid is described by the linearized Navier-Stokes equation. The profile of the calculated transverse displacement is similar to that measured by \citet{Tasaki1988}. A coupled model of electrical and mechanical signals based spring-dampers (dashpots) system has been elaborated by \citet{Jerusalem2014}. The ion currents are calculated again using the HH model and calibrated for a guinea pig spinal cord white matter. This model provides a framework for damage mechanisms in neurons. For this purpose a special simulation package Neurite has been developed \citep{Garc??a-Grajales2015}.

As far as the governing wave equations modelling of all the single processes have actually been derived, the challenge is to formulate a model based on the system of coupled governing equations. First ideas on such a model are described by \citep{Engelbrecht2016}. Further this model is elaborated in more detail.

\section{A model involving an ensemble of waves}

In general terms, beside electrophysiology and mechanisms in biomembranes, the ideas from continuum theory area used (see also \citet{Lomholt2006}). The general concepts well-known in mathematical physics are followed – the initial conditions and forcing are formulated in variables involved in governing equations.  

The starting assumptions in modelling are the following:\\
(i) electrical signals are the carriers of information \citep{Debanne2011} and trigger all the other processes;\\
(ii) the axoplasm in a fibre can be modelled as a viscous fluid where a pressure wave is generated due to electrical signal \citep{Terakawa1985,Rvachev2010,Hady2015};\\
(iii) the biomembrane can be deformed \citep{Gross1983,Heimburg2005} in the longitudinal as well as in the transverse direction;\\
(iv) the channels in biomembranes can be opened and closed under the influence of electrical signals as well as of the mechanical input \citep{Heimburg2010,Mueller2014}.\\
The aim is to use known mathematical models (governing equations) but adding the contact forces which need additional assumptions. The first approach described below is to build up as simple (robust) system as possible in order to test assumptions, especially on coupling forces.

The process is initialized by an electrical input $f(z)$ which is 
\begin{equation}
\label{initpulse}
z|_{t=0}=f(x),
\end{equation}
\begin{wrapfigure}{l}{0.25\textwidth}
 \includegraphics[width=0.25\textwidth]{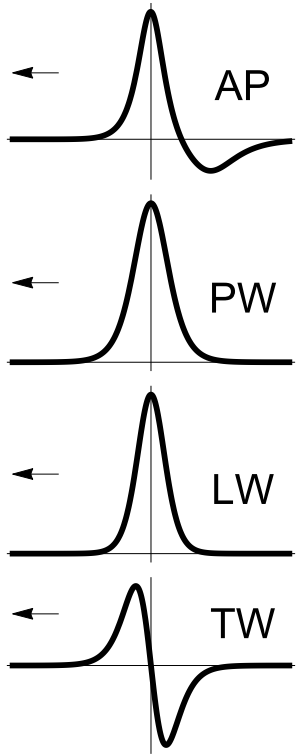}
  \caption{Schemes of the ensemble of waves. Here AP - action potential, PW - pressure wave in axoplasm,  LW - longitudinal wave in the biomembrane (BM), TW - transverse wave in the BM, scales are arbitrary. Reproduced from \citet{Engelbrecht2018}.}
\label{Waveschemes}
\end{wrapfigure}
where $z$ is an electrical pulse above the threshold level. The action potential AP is governed by a FHN-type model \citep{Nagumo1962} in the form of two coupled equations:
\begin{align}
\label{FHN1}
	&z_t=z(z-(a_1+b_1))(1-z)-j+D \, z_{xx},\\
	\label{FHN2}
	&j_t=\varepsilon(-j+(a_2+b_2)z),
\end{align}
where $z$ is a scaled voltage, $j$ is the recovery current, $D$ is a coefficient, $\varepsilon$ is the time-scale difference (see \citet{Nagumo1962}) and $0<a_1+b_1<1$, $a_2+b_2>0$, $x$ and $t$ are dimensionless space and time respectively. Here $a_1$, $a_2$ control the `electrical' activation and added coefficients $b_1$, $b_2$ control the `mechanical' activation.\\
The pressure wave is governed by Eq.~\eqref{LinMorganWaveEQ} with a driving force
\begin{equation}
	\label{LinMorganPressureWaveForced}
	\bar{p}_{tt}=c_f^2\bar{p}_{xx}  - \mu \bar{p}_{t} +F_1(z,j),
	\end{equation}
where $F_1(z,j)$ is a force from the AP and $\mu \bar{p}_t$ is added viscous dampening term. At this moment we leave open whether the changes in the voltage or in the ion current play role of a  driving force. %It is possible to include also the viscosity $-\mu\bar{p}_t$ into Eq.~\eqref{LinMorganPressureWaveForced}. \\
In the biomembrane, the governing equation for a longitudinal wave is derived from the balance of momentum with the special ‘displacement-type’ nonlinearity and dispersive terms \citep{Heimburg2005,Engelbrecht2015}:
\begin{equation}
\label{HJimproved}
u_{tt}=\left[\left(c_0^2+pu+qu^2\right)u_{x}\right]_{x}-h_1 u_{xxxx} + h_2 u_{xxtt} +F_2(j,\bar{p}),
\end{equation}
where $u=\Delta\rho_0$ is the density change of a biomembrane, $c_0$ is the velocity in the unperturbed state, $p$, $q$ are coefficients of nonlinearity, $h_1$, $h_2$ are dispersion constants and $F_2(j,\bar{p})$ is the force exerted by the processes in the axoplasm. \\
Finally, the transverse wave w, following the ideas from the theory of rods \citep{Porubov2003} is governed by 
\begin{equation}
\label{transversedispl}
w=-kr \cdot u_{x},
\end{equation}

where $r$ is the radius of the fibre and $k$ is a constant. In the theory of rods $k$ is the Poisson ratio.

Some remarks concerning equations \eqref{initpulse}-\eqref{transversedispl} are in order. The AP is described by a simple FHN model involving only one ion current \citep{Nagumo1962}. One could certainly use the HH model with two (sodium and potassium) ion currents \citep{Hodgkin1945,Hodgkin1964a} or even a generalized model with more ion currents \citep{Courtemanche1998} but with the aim to test the coupling forces, we start with this robust simpler model. The limitation is that the effects of anaesthesia are oversimplified. The pressure wave could certainly be described also by a 2D Navier-Stokes model. This change must be considered with a special attention because if the transverse velocity $v_y$ will be taken into account, it could modify the forces exerted to the biomembrane. It must also be stressed that in the improved model describing longitudinal waves in a biomembrane \citep{Engelbrecht2015}, the second dispersive term with the coefficient $h_2$ describes the microinertia of the lipid bilayer and corresponds to principles of continuum theory for microstructured solids \citep{Engelbrecht2005}.

\begin{wrapfigure}{r}{0.52\textwidth}
\vspace{-0.7cm}
	\centering
	\includegraphics[width=0.51\textwidth]{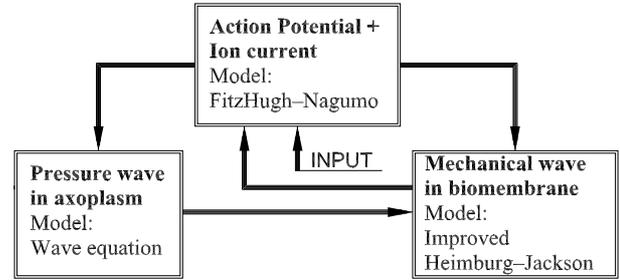}
	\caption{Block diagram of the combined model for the nerve pulse propagation.}
	%\vspace{-1.5cm}
	\label{blokkd}
%\end{figure}
\end{wrapfigure}

As a result we expect an ensemble of waves to be generated which is schematically shown in Fig.~\ref{Waveschemes} and a block diagram depicting the relationships between the individual components of the proposed model is shown in Fig.~\ref{blokkd}.

\section{Results of numerical simulation}
%\begin{figure}
\begin{wrapfigure}{l}{0.6\textwidth}
\centering
\includegraphics[width=0.51\textwidth]{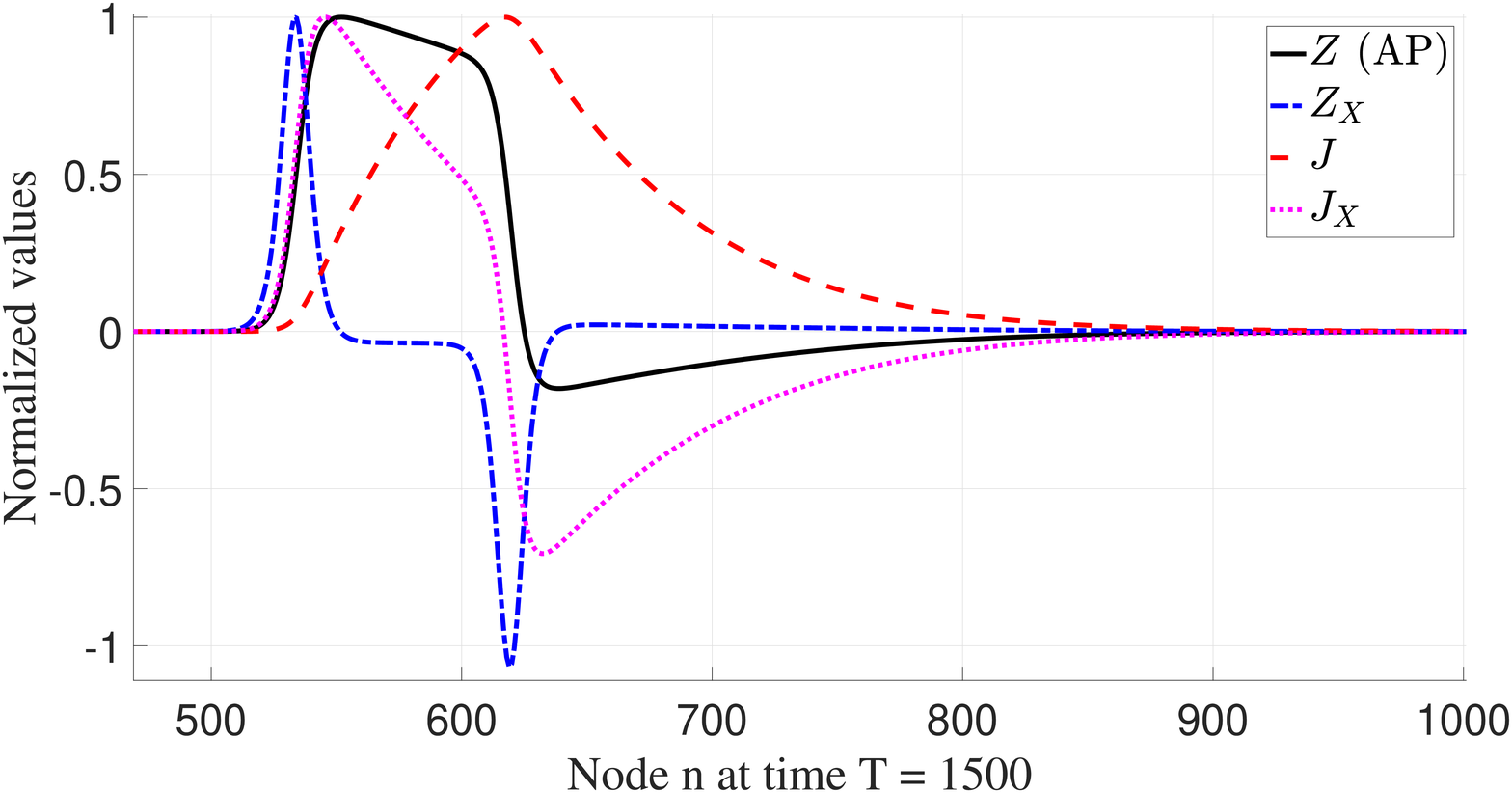}\\ 
\vspace{4mm}
\includegraphics[width=0.51\textwidth]{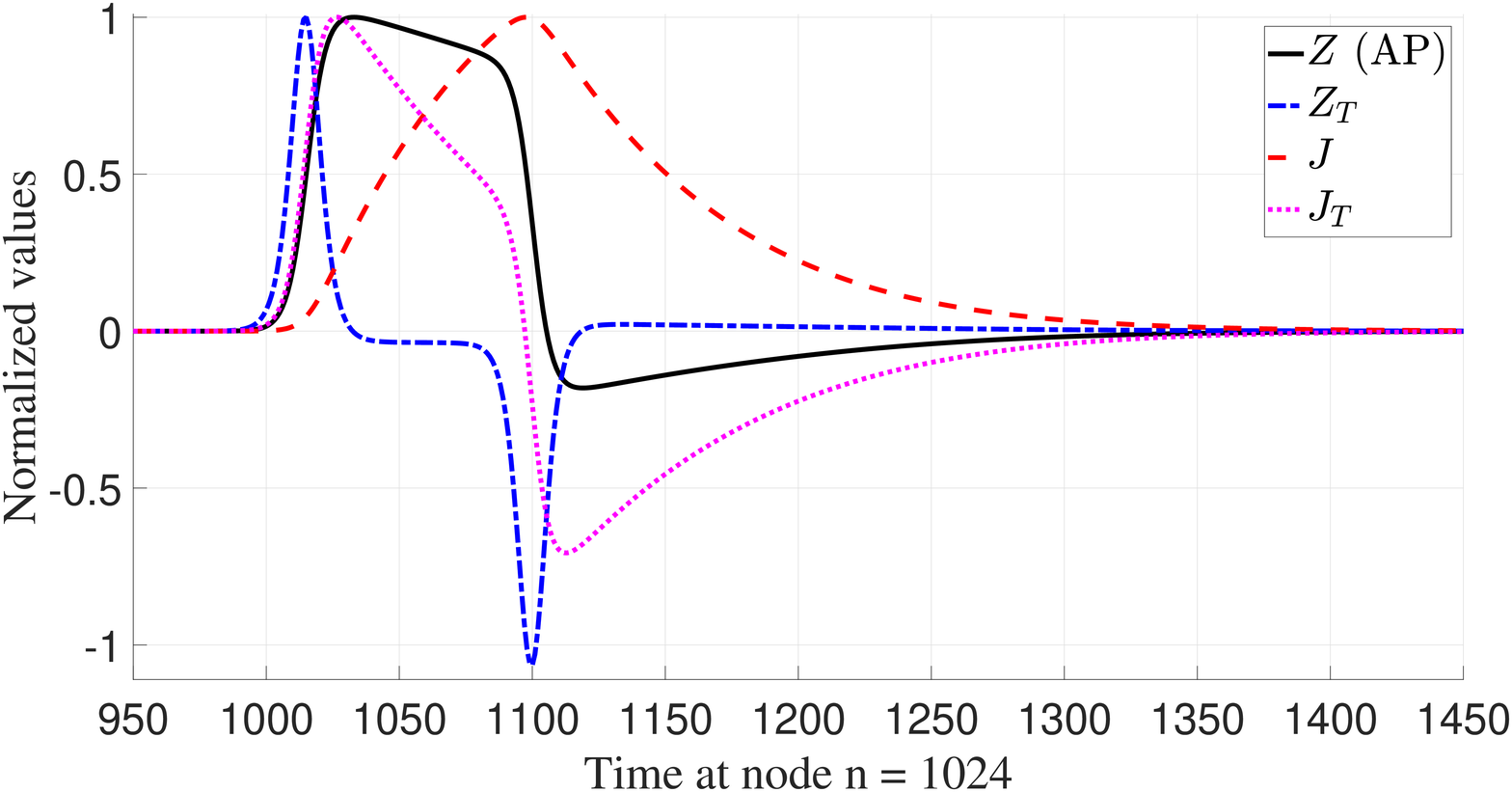}
\caption{The solutions and their derivatives of the FHN equation. Top panel -- action potential $Z$, recovery current $J$ and their gradients $Z_X, J_X$ in space at $T=1500$, bottom panel -- action potential $Z$, recovery current $J$ and their time derivatives $Z_T, J_T$ in time at spatial node $n=1024$.} 
\label{ioncurrent}
%\end{figure}
\end{wrapfigure}

The most important problem in building the joint model is related to the assumptions about the coupling forces. Although the various mechanisms of transduction between fields are analyzed \citep{Gross1983,Gonzalez-Perez2016}, there is still no widely accepted understanding about the character of this process. Here we follow an assumption that the mechanical waves are generated by two changes in electrical pulses: either in the AP or in the ion current \citep{Engelbrecht2018} and by the changes in pressure in the axoplasm. In more general terms this means that the dynamical processes are not generated by values of the fields but by changes in the field. Consequently, we assume that 
\begin{align}
	\label{AssumptionCouplingConstants}
	&F_1=\eta_1z_x+\eta_2 j_t,\\
	&F_2=\gamma_1\bar{p}_t+\gamma_2j_t,
\end{align}
where $\eta_1$, $\eta_2$, $\gamma_1$, $\gamma_2$ are suitable coefficients. Further on, the normalized values of variables are used in calculations ($Z$ for the AP amplitude, $J$ for the ion currents, $\bar{P}$ for the pressure, $U$ for the LW amplitude) together with dimensionless space and time coordinates $X$, $T$. The normalization of independent variables is based on Eq.~\eqref{HJimproved} where the velocity $c_0$ and the characteristic length of an axon is used \citep{Engelbrecht2015}. For example, the generated ion current calculated from the FHN model and its gradients $Z_X$, $I_X$ as well as $Z_T$, $J_T$ are shown in Fig.~\ref{ioncurrent}. In principle, the bi-polarity is evident for all derivatives.  

Note that the exact nature of the coupling forces terms is left open at this stage. One possible physical interpretation of the proposed terms is that the time derivatives could be interpreted as forces acting across the lipid bi-layer at a fixed spatial point on axon while spatial gradients could be interpreted as forces acting along the axon axis. For example, the force $F_1$ in the pressure expression could contain two terms -- $Z_Z$, $J_T$. First, an action potential gradient $Z_X$ could be related to the fact that there are charged particles (ions) present inside the axon that might move along the axis of the axon in the presence of the potential gradient. Second, an ion current time derivative $J_T$ could be related to changes in pressure when the ions flow in and out of the axon through the lipid membrane during the nerve pulse propagation at a fixed point (ion channel) on the axon. As noted earlier we use a simplified model for the action potential where all the ion currents present are wrapped into one abstracted ion current but if one would be using one of the more complex models (HH model, for example) the similar logic can be extended to any number of individual ion flows and to include some parameters specific to ion channel behaviour for these individual ion flows.
%\begin{figure} 

The assumption on $j_x$ or $j_t$ (see Fig.~\ref{ioncurrent}) as a driving force has an important property – the force exerted to the biomembrane is bipolar and is therefore energetically balanced. If a localized pulse-type force is used then the energetical balance due to the moving signal $z(x,t)$ is distorted by the continuous energy influx. 
The next question is related to the composition of an ensemble and the relative significance of all three constituents in it: the electrical signal, the pressure wave in the axoplasm and the mechanical wave in the biomembrane. The measured pressure change in the axoplasm is extremely small \citep{Terakawa1985}.  The transverse displacements of the fibre wall (biomembrane) are also small but can be measured \citep{Tasaki1988}.

As far as the coupling mechanisms of electrical and mechanical signals are not fully understood, we shall use mathematical simulation with the goal to understand how the coupling process can be modelled in terms of the governing equations of single waves. Before simulation of three waves (AP, PW, LW), we proceed with simpler two-wave models: AP and LW coupled and AP and PW coupled. All the numerical calculations are carried out by using the pseudospectral method (see Appendix).

The system of model equations solved numerically in the dimensionless forms is
\begin{equation}
\begin{split}
& Z_{T} = D Z_{XX} + Z \left( Z - \left[ a_1 + b_1 \right] - Z^2 + \left[ a_1 + b_1 \right] Z \right) - J, \\
& J_{T} = \varepsilon \left( \left[ a_2 + b_2 \right] Z - J \right),\\
& U_{TT} = c^2 U_{XX} + P U U_{XX} + Q U^2 U_{XX} + P U_{X}^{2} + 2 Q U U_{X}^{2} - H_1 U_{XXXX} + H_2 U_{XXTT} + \gamma_1 \bar{P}_T + \gamma_2 J_T,\\
& \bar{P}_{TT} = c_{f}^{2} \bar{P}_{XX} - \mu \bar{P}_T + \eta_1 Z_X + \eta_2 J_T ,
\end{split}
\label{EQS1}
\end{equation}
where capital letters denoting dependent variables are used to emphasize that we are dealing with the dimensionless case. As noted earlier $Z$ is the action potential, $J$ is the recovery current, $a_i, b_i$ are the `electrical' and `mechanical' activation coefficients, $D, \varepsilon$ are coefficients, $U$ is the longitudinal density change in lipid layer, $c$ is the velocity of unperturbed state in lipid bi-layer, $P, Q$ are the nonlinear coefficients, $H_1, H_2$ are the dispersion coefficients and $\gamma_1, \gamma_2$ are the coupling coefficients for the mechanical wave, $\bar{P}$ is the pressure, $c_{f}$ is the characteristic velocity in the fluid, $\eta_1, \eta_2$ are the coupling coefficients for the pressure wave and $\mu$ is the (viscous) dampening coefficient. `Mechanical' activation coefficient could be connected to the improved Heimburg-Jackson model as $b_1 = - \beta_1 U$ and $b_2 = - \beta_2 U$ where $\beta_1, \beta_2$ are the mechanical coupling coefficients which could be different for the action potential and recovery current parts of the FHN equations. Note that in system~\eqref{EQS1} either $J_T$ or $J_X$ can be used as coupling forces, the coupling terms used for a given calculation are noted in the figure legends after the variable plotted. The localized initial conditions and periodic boundary conditions are used (see Appendix for details on the numerical scheme). The coupling coefficients vary and the rest of the parameters are the same for all the depicted solutions. The common parameter values are used: $D=1, \varepsilon=0.01, a_1=0.2, a_2=0.2, c^{2}=0.16, P=-0.05, \break Q=0.02, H_1= 0.43, H_2=0.75, c_{f}^{2}=0.1, \mu=0.0025$

%\begin{figure}
\begin{wrapfigure}{R}{0.52\textwidth}
\centering
\includegraphics[width=0.51\textwidth]{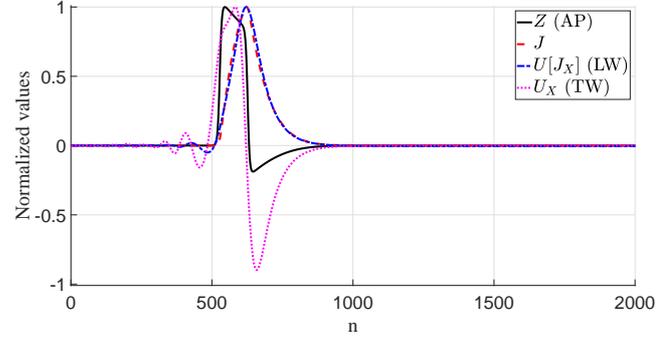}
\caption{Action potential coupled with the mechanical wave. Solutions at $T=1500$. The coupling parameters are $\beta_1=\beta_2=0.05, \break \gamma_1=0, \gamma_2=0.002, \eta_1=\eta_2 =0$.} 
\label{Fig4}
%\label{joonis1}
%\end{figure}
\end{wrapfigure}
%
%\begin{figure}

\paragraph{(i) Two-wave model I} In this case we neglect the pressure wave (PW) in the axoplasm and formulate a model including the electrical signal (AP) in the fibre and the accompanying longitudinal wave (LW) in the biomembrane. Then the coupled model includes Eqs~\eqref{FHN1}, \eqref{FHN2} and Eq.~\eqref{HJimproved}. In the latter, the force $F_2(j, \bar{p})$ is taken as $F_2(j)$ only, i.e., depending only on the AP. The detailed analysis of this case is presented by \citet{Engelbrecht2018}. 
In terms of system~\eqref{EQS1} this means that $\gamma_1=\eta_1=\eta_2=0$.
The main features are the following: \\
%\begin{itemize}
	%\item[\textendash] 
	-- the input (the initial condition) for Eqs~\eqref{FHN1}, \eqref{FHN2} is taken as a narrow 
	$\sech^2$–type pulse with an amplitude above the threshold; \\  
	%\item[\textendash] 
	-- the generated electrical pulse (AP) has a typical asymmetric form with an overshoot and generates an ion current; \\
	%\item[\textendash] 
	-- the gradient (i.e., the change) of the ion current is taken as an input for the generation of the mechanical longitudinal wave (LW); \\
	%\item[\textendash] 
	-- the derivative of the LW gives the profile of the TW \citep{Engelbrecht2015}. \\
%\end{itemize}
Note that (i) the gradient of the ion current is energetically balanced; (ii) the velocities of the AP and LW are chosen to be synchronized.

The simulation results in the dimensionless form are shown in Fig.~\ref{Fig4} which demonstrates the profiles of the AP, LW and TW together with the ion current. The latter has a characteristic shape measured by \citet{Iwasa1980,Tasaki1988} and \citet{Gonzalez-Perez2016}.

\begin{wrapfigure}{R}{0.52\textwidth}
\centering
\includegraphics[width=0.51\textwidth]{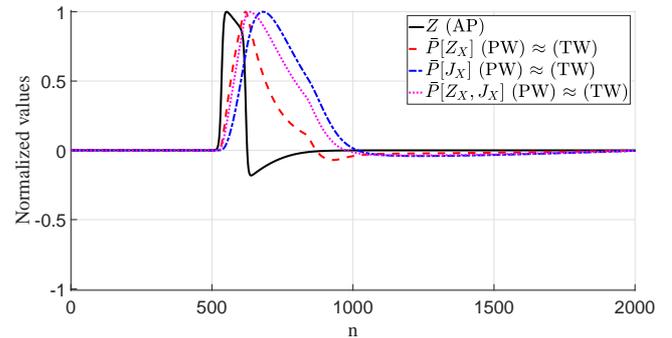}
\caption{Action potential coupled with the pressure wave (two different coupling forces considered). Solutions at $T=1500$. The coupling parameters are $\beta_1=\beta_2=0.0, \gamma_1=\gamma_2=0.0, \eta =0.002 (Z_X),\break \eta =0.02 (J_X)$. In the case of $\bar{P}[Z_X,J_X]$ the coupling parameters are $\eta_1=0.001 (Z_X), \eta_2=0.01 (J_X)$.} 
\label{FigAPandPW}
%\end{figure}
\end{wrapfigure}

\paragraph{(ii) Two-wave model II} In this case we formulate a model in terms the electrical signal AP and the pressure wave PW. The model involves Eqs~\eqref{FHN1}, \eqref{FHN2} and governing equation \eqref{LinMorganPressureWaveForced} for the pressure. In terms of system~\eqref{EQS1} this means that $\beta_1=\beta_2=\gamma_1=\gamma_2=0$.

\begin{wrapfigure}{r}{0.52\textwidth}
	\centering
	\includegraphics[width=0.51\textwidth]{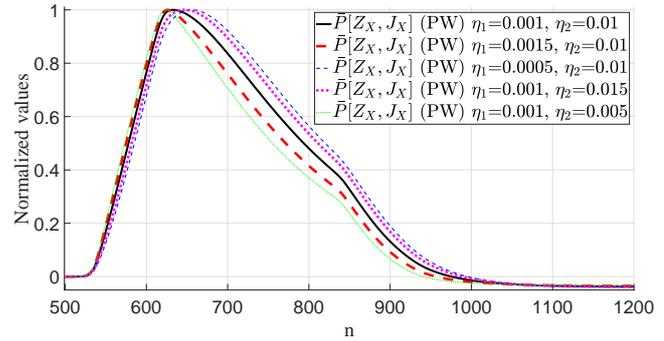}
	\caption{Pressure wave profiles with different coupling parameters at $T=1500$.}
	\label{DifferentPWProfiles}
	%\end{figure}
\end{wrapfigure}

\begin{figure}[h]
	\centering
	\includegraphics[width=0.95\textwidth]{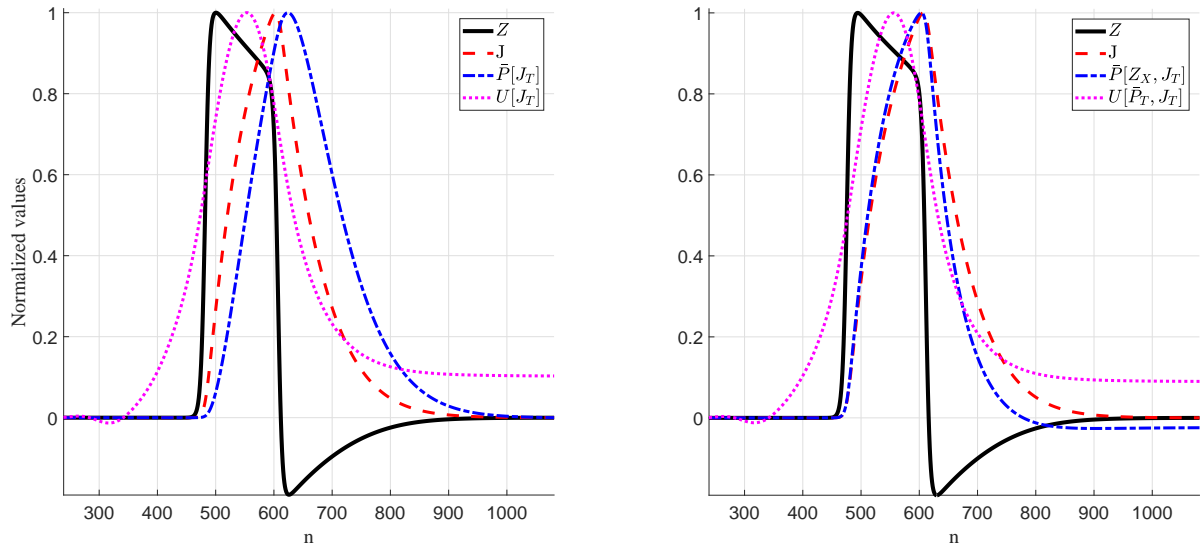}
	\caption{The solutions of the three wave model at $T=1500$ when using the ion current time derivative $J_T$ (left panel) and $J_T$ plus pressure time derivative $P_T$ and action potential gradient $Z_X$ as a coupling forces (right panel). Parameters $\beta_1=\beta_2=0.05,\break \gamma_1=0, \gamma_2=0.01, \eta_1=0, \eta_2=0.01$ (left panel) and  $\beta_1=\beta_2=0.05, \gamma_1=0.001, \gamma_2=0.01, \eta_1=0.001, \eta_2=0.01$ (right panel).}
	\label{FigJt}
\end{figure}

\begin{wrapfigure}{R}{0.49\textwidth}
\centering
\includegraphics[width=0.48\textwidth]{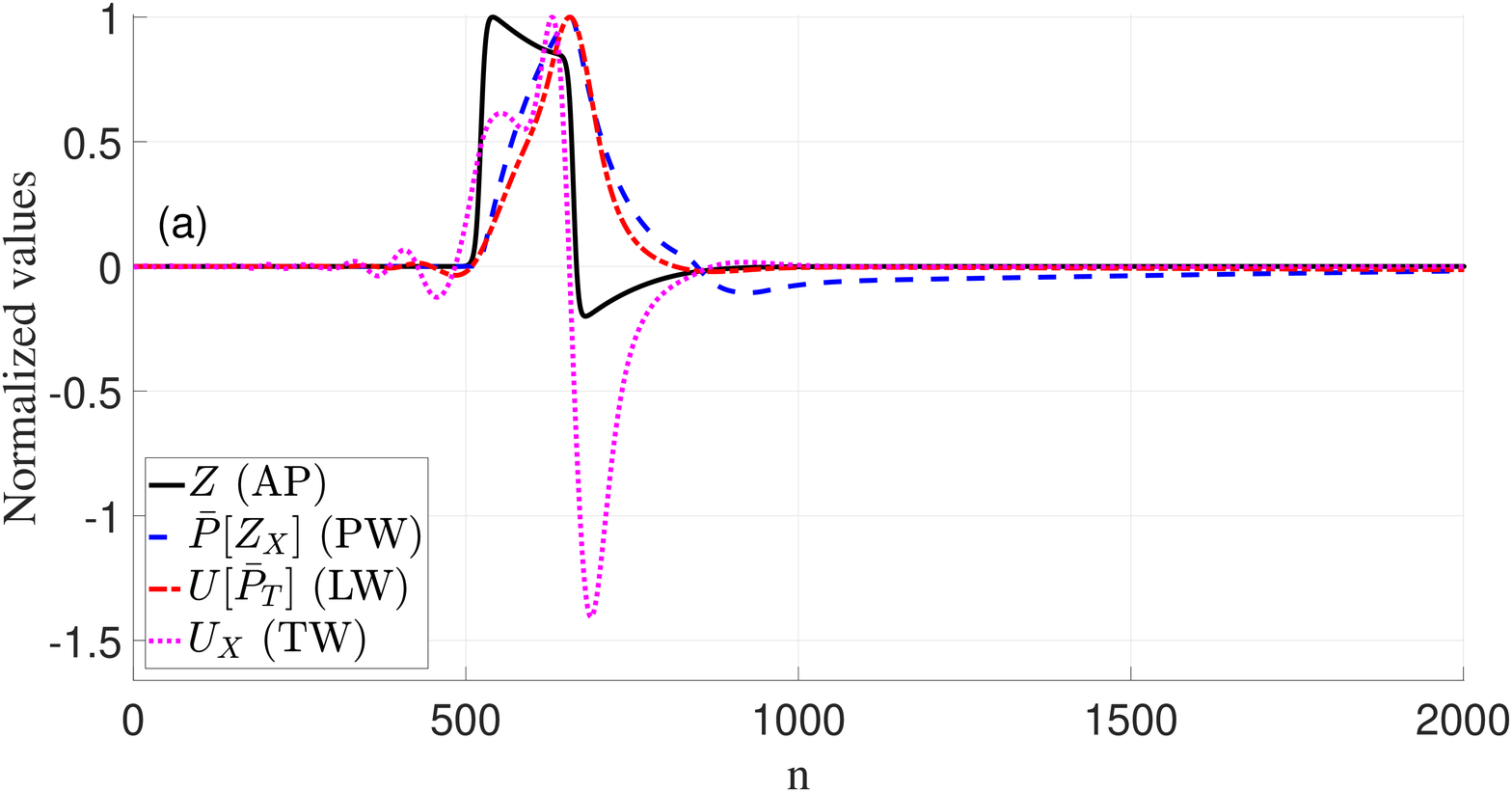}\\
\vspace{2mm}
\includegraphics[width=0.48\textwidth]{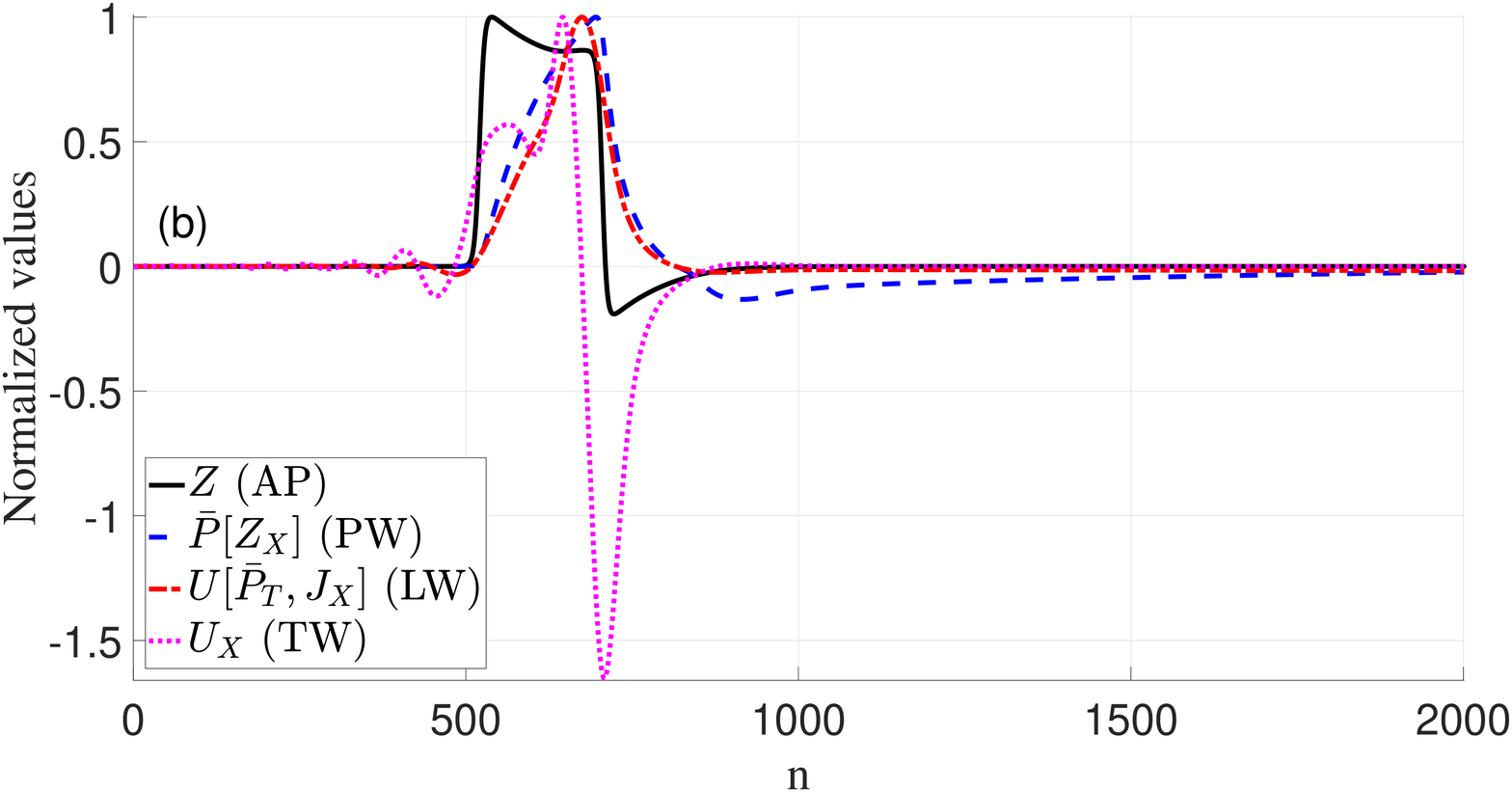}\\
\vspace{1.8mm}
\includegraphics[width=0.48\textwidth]{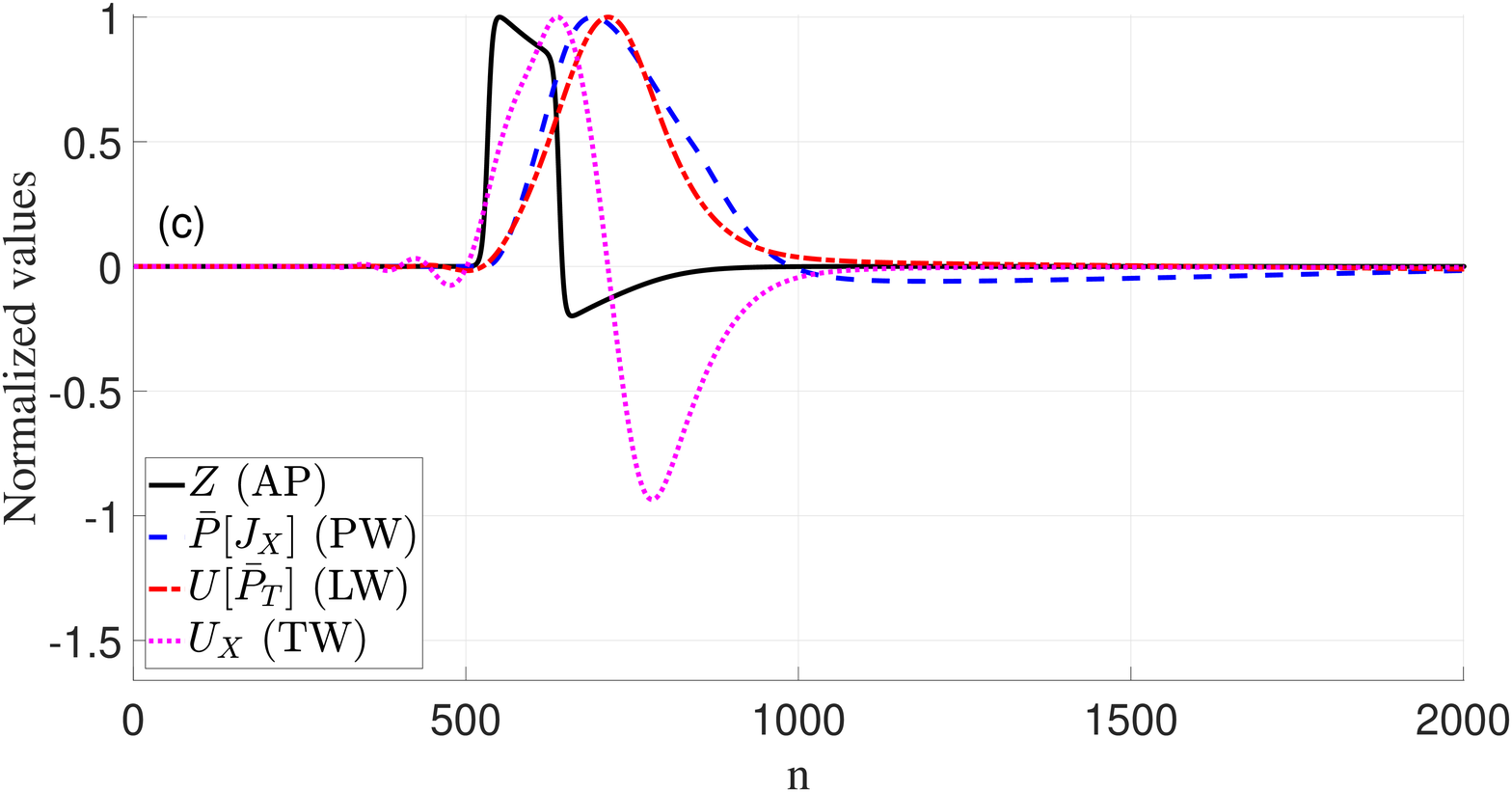}\\
\vspace{1.8mm}
\includegraphics[width=0.48\textwidth]{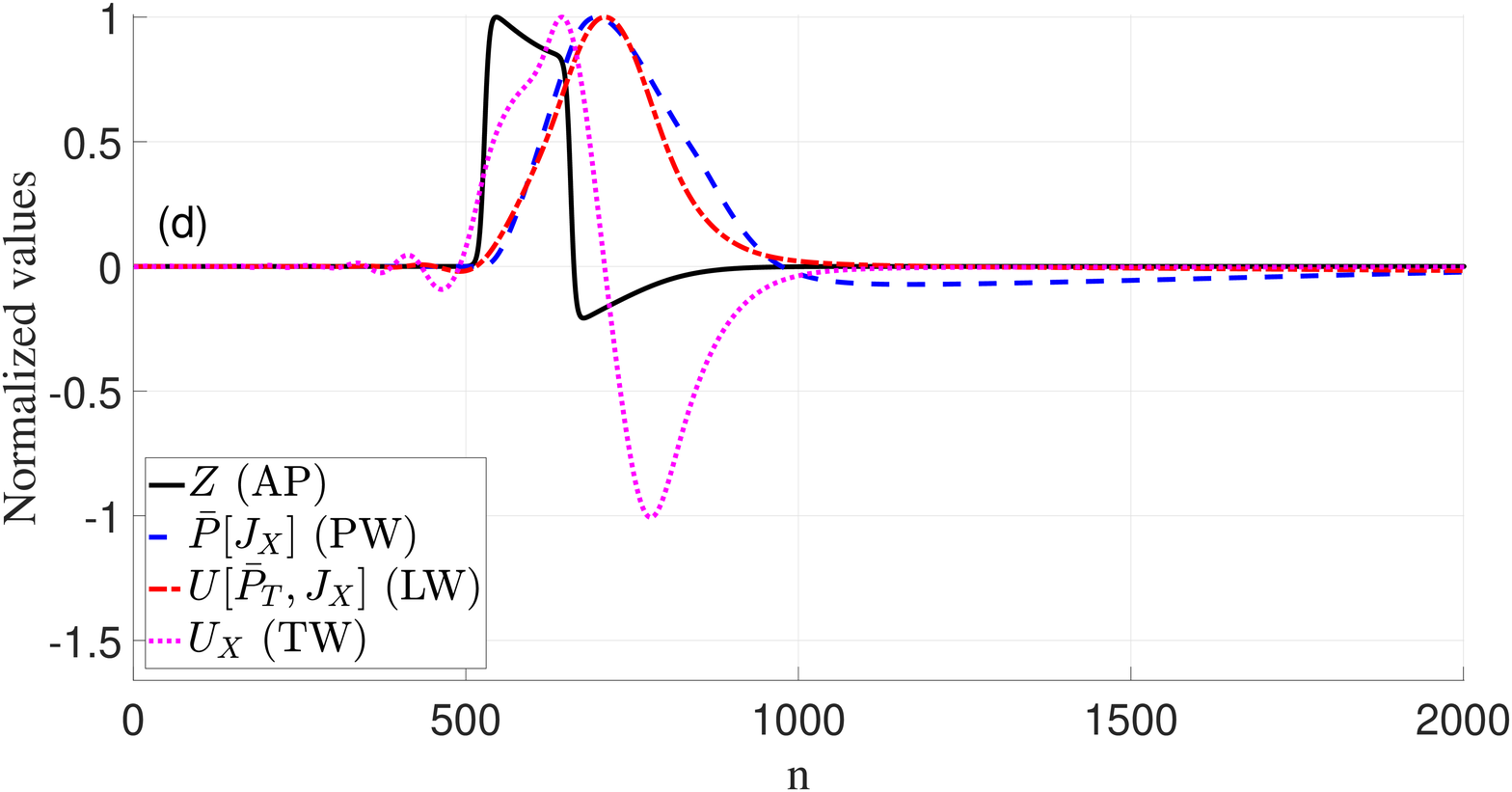}\\
\vspace{1.8mm}
\includegraphics[width=0.48\textwidth]{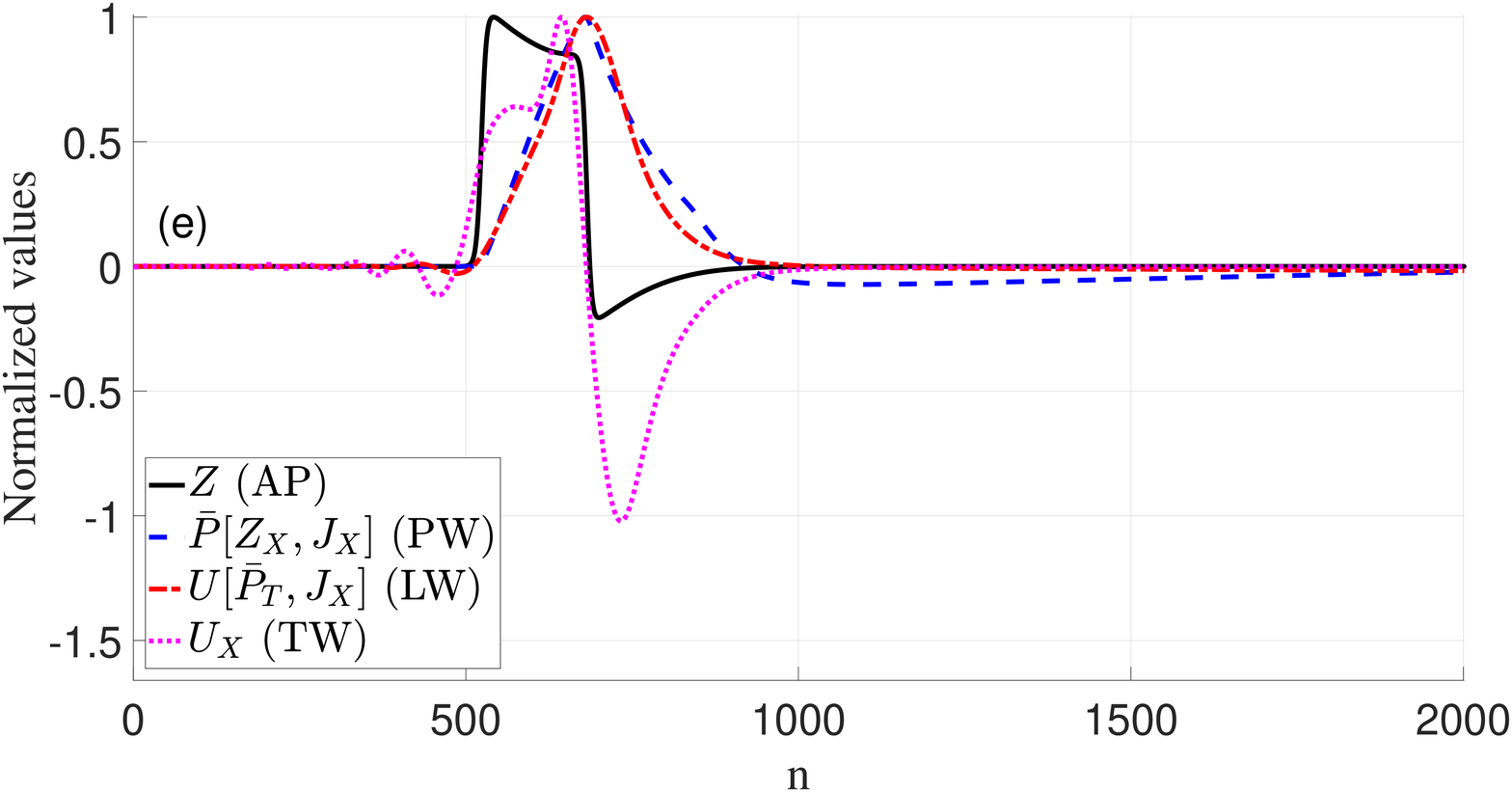}
\caption{The solutions of the three wave model when using the ion current gradient $J_X$ as one of the coupling forces. See text for parameter details. } 
\vspace{-9mm}
\label{ThreeCompFig}
%\end{figure}
\end{wrapfigure}

The simulation results are shown in Fig.~\ref{FigAPandPW} and the pressure profiles for different combinations of coupling parameters $\eta_1$ and $\eta_2$ are shown in Fig.~\ref{DifferentPWProfiles}. The pressure wave (PW) modelled by Eq.~\eqref{LinMorganPressureWaveForced} demonstrates retardation from the AP and a slight overshoot \citep{Terakawa1985}. As far as the wave equation \eqref{LinMorganPressureWaveForced} has pretty stable solutions, the small changes is the coefficients $\eta_1$, $\eta_2$ which characterize the driving force $F_1$, do not lead to essential changes in the profile of the PW (see Fig.~\ref{DifferentPWProfiles}). Increasing $\eta_1$ leads to a steeper front and faster decay at the back of the profile, while the effect of the $\eta_2$ is opposite.
\newpage
	%\begin{figure}
\paragraph{(iii) Three-wave model} In this case all three components of a signal – AP, PW, LW – are taken into account. An important question is to estimate the forms of physically plausible contact forces $F_1$ and $F_2$. From the analysis of case (i) with coupled AP and LW it is possible to conclude that the character of the $F_2$ should be bipolar. The numerical simulation permits to calculate the profiles with several forces depending on $Z_X$, $P_T$, $J_X$, $J_T$. The corresponding wave profiles at $T=1500$ are shown in Fig.~\ref{FigJt} for the case when time derivatives are used for coupling forces and in Fig.~\ref{ThreeCompFig} for the cases where mostly gradients are used as coupling terms. 
In Fig.~\ref{ThreeCompFig}: \\
(a) -- The pressure wave is generated by the action potential gradient and the mechanical wave is generated by the pressure time derivative. The coupling parameters are $\beta_1=\beta_2=0.05, \gamma_1=0.002, \gamma_2=0,\break \eta_1 =0.002, \eta_2=0$.\\
(b) -- The pressure wave is generated by the action potential gradient and the mechanical wave is generated by the pressure time derivative and the ion current gradient. The coupling parameters are\break  $\beta_1=\beta_2=0.05, \gamma_1=\gamma_2=0.002, \eta_1 =0.002, \eta_2=0$.\\
(c) -- The pressure wave is generated by the ion current gradient and the mechanical wave is generated by the pressure time derivative. The coupling parameters are $\beta_1=\beta_2=0.05, \gamma_1=0.002, \gamma_2=0, \eta_1=0, \eta_2 =0.02$.\\
(d) -- The pressure wave is generated by the ion current gradient and the mechanical wave is generated by the pressure time derivative and ion current gradient. The coupling parameters are $\beta_1=\beta_2=0.05, \break\gamma_1=\gamma_2=0.002, \eta_1=0, \eta_2 =0.02$.\\
(e) -- The pressure wave is generated by the ion current gradient and action potential gradient while the mechanical wave is generated by the pressure time derivative and ion current gradient. The coupling parameters are $\beta_1=\beta_2=0.05, \gamma_1=\gamma_2=0.002, \eta_1 =0.001, \eta_2 =0.01$.

From the viewpoint of the behaviour of the solution there is almost no qualitative difference if we use a time derivative or spatial gradient as the coupling term because as demonstrated in Fig.~\ref{ioncurrent}, the shape of the function is the same in essence. In the used numerical scheme the calculation of spatial derivatives is more convenient and numerically more accurate and for that reason in the following analysis the focus is on the case where $J_X$ is used as one of the coupling terms. Numerically we find $J_X$ by making use of the properties of the fast Fourier transform while for finding $J_T$ a simple backward difference scheme is used. However, if the experiments demonstrate the need to use $J_T$ in coupling forces, this can also be realized. 

The profiles in Figs~\ref{FigJt} and~\ref{ThreeCompFig} demonstrate a typical AP with an overshoot, a pressure wave (PW) propagating behind the AP and the longitudinal wave (LW) in the biomembrane with a typical solitary wave profile. Feedback coupling is taken into account for the AP from the LW and its influence is more evident in Figs~\ref{ThreeCompFig}b, \ref{ThreeCompFig}e. These profiles correspond qualitatively to previous studies starting from the AP \citep{Hodgkin1945,Nagumo1962} to experimentally measured PW \citep{Terakawa1985} and LW \citep{Heimburg2005,Gonzalez-Perez2016}. The transverse wave TW is calculated from the LW by using expression \eqref{transversedispl} and has a bipolar shape \citep{Tasaki1988}. Note that all the profiles are dimensionless with their maximal amplitude taken as a scaling measure.

The basic assumption in all calculations is that the coupling is influenced by the changes of the field quantities, not by their values. This idea is supported by several studies \citep{Terakawa1985,Kim2007,Mueller2014,Gonzalez-Perez2016}. The initial stage of the AP forming from input \eqref{initpulse} is not analyzed because of possible fast changes and presented analysis takes a fully formed AP as a basic signal for coupled waves. The profiles in Figs~\ref{FigJt} and \ref{ThreeCompFig} are qualitatively similar to all measured ones. The parameters for simulation shown in Figs~\ref{FigJt} and \ref{ThreeCompFig} have been chosen to generate mechanical effects a little bit behind the AP. This brings up the question about the synchronization of velocities. In principle, the wave velocities in continua depend on elastic properties and density but due to coupling effects the group velocities (responsible for energy propagation) may considerably differ from the sound velocity due to dispersion. The velocity of the AP may also be affected by axonal irregularities and ion channels \citep{Debanne2011}. Note also that the velocity of the blood flow in a vessel depends on the stiffness of the vessel wall \citep{Barz2013}. So, we have to agree that ``the conduction velocity of mechanical impulses in nerve fibres is unknown'' \citep{Barz2013} and needs further theoretical and experimental studies in order to establish joint understanding. So, as a proof of concept, this study has fulfilled the idea to look for the biomembrane mediated signalling in a nerve fibre as a complex system, resulting in an ensemble of waves. 

It can be concluded from profiles shown in Figs~\ref{FigJt} and~\ref{ThreeCompFig} that the influence of changes for coupling can be presented either by gradients $Z_X$, $J_X$ or by the time derivatives $P_T, J_T$  or in the more general case as some combination of the considered coupling terms.

\section{Discussion}

Clearly the analysis carried out above is only the first stage of modelling. The AP is calculated by a simple FHN model with only one ion current but the HH-model includes both sodium and potassium currents. In this case the number of coefficients is certainly much higher \citep{Courtemanche1998} but gives a lot of possibilities to model anaesthetics \citep{Heimburg2007}. In addition, the time shift of sodium and potassium ion currents may give an additional possibility to specify the generation on mechanical effects. The more detailed handling of the ion currents would certainly enable accounting for the effect of individual ion flows when coupled to the equations governing the pressure and mechanical waves. One could for example consider the effect of the ion sizes, charges and masses.

Temperature effects and possible heat production \citep{Heimburg2010} are also not taken into account. It is also discussed whether water movement across the biomembrane associated with sodium influx may affect the mechanical effects \citep{Kim2007}. For the pressure wave as a first approximation the wave equation with added dampening term is adequate for catching the main effect of a disturbance propagating in the viscous environment. The obvious direction for improvement would be the celebrated Navier-Stokes equations allowing to account for compressibility, non-linearity, viscosity, etc. from the first principles. 
\newpage %\vspace{-3mm}
%\begin{figure}[H]
\noindent
Also, if a HH-like model for the AP is used, then the effects of individual ion currents on the pressure could be studied in greater detail. So there are several possibilities to improve the mathematical models.

Even when using component models with higher level of detail the qualitative picture should remain similar. There might emerge some finer nuances that the simpler models cannot quite catch in full detail. However, the basic principle of being able to combine the components into a whole which is richer than just the sum of individual components through the coupling forces will still hold. As noted in the Introduction, the scientists are good at breaking the complex problems down into simpler problems which can be solved that they sometimes forget to put things back together (Toffler, 1984) -- this is one possibility of putting different aspects of such a complicated phenomenon as the nerve pulse propagation back together.

The former studies \citep{Jerusalem2014,Hady2015} have pointed out several possibilities to build up models which could describe a coupled signal. To the best knowledge of authors, the model presented above is the first attempt to compose the governing differential equations of single waves into a system coupled by interaction forces resulting in an ensemble of waves. It is, as stated above, a proof of concept in terms of mathematical physics. The model needs certainly experimental verification. 
Compared with the classical experiments \citep[etc.]{Terakawa1985,Tasaki1988} there are now contemporary powerful experimental methods like atomic force microscopy \citep{Kim2007}, optical detectors \citep{Perez-Camacho2017}, and other methods described in many studies \citep[see][etc.]{Clay2005,Scholkmann2014,Gonzalez-Perez2016}. The next decade will surely bring along exciting results in measuring the ensemble of waves.

Finally, as stated by \citet{Kaufmann2018} in his analysis of paradoxes in physics: ``the origin of the nervous impulse unifies the realities'' referring to studies of Hill-Hodgkin-Tasaki. Indeed, already \citet{Hill1936} has shown the importance of ion currents, \citet{Hodgkin1964a} called for ``a tidy pattern'' and \citet{Tasaki1988} explained the non-electrical manifestation of excitation process. It is a challenge to incorporate all observed phenomena in one theory \citep{Andersen2009}. The present paper proposes a robust mathematical explanation to the coupling of waves in nerve fibres. We followed the principle of Ockham's razor which states simply that more things should not be used than are necessary. However, we admit that given the complicated structure of cells, the coupling forces may have much more complicated structure than proposed within this model.

\section*{Acknowledgements} 
 
 This research was supported by the European Union through the European Regional Development Fund (Estonian
Programme TK 124) and by the Estonian Research Council (projects IUT 33-24, PUT 434).

\begin{appendices}
\setcounter{equation}{0}
\numberwithin{equation}{section}
\section{The numerical scheme}
\subsection{The system of partial differential equations to be solved numerically}
As noted above, the pseudospectral method (PSM) (see \cite{Fornberg1998,Salupere2009}) is used to solve the system of dimensionless model equations:
\begin{equation}
\begin{split}
& Z_{T} = D Z_{XX} + Z \left( Z - \left[ a_1 + b_1 \right] - Z^2 + \left[ a_1 + b_1 \right] Z \right) - J, \\
& J_{T} = \varepsilon \left( \left[ a_2 + b_2 \right] Z - J \right),\\
& U_{TT} = c^2 U_{XX} + P U U_{XX} + Q U^2 U_{XX} + P U_{X}^{2} + 2 Q U U_{X}^{2} - H_1 U_{XXXX} + H_2 U_{XXTT} + \gamma_1 \bar{P}_T + \gamma_2 J_T,\\
& \bar{P}_{TT} = c_{f}^{2} \bar{P}_{XX}  - \mu \bar{P}_T + \eta_1 Z_X + \eta_2 J_T.
\end{split}
\label{EQS}
\end{equation}
The notations used here are already given in the main text.
%
% but is repeated here for the sake of completeness. Here $Z$ is the action potential, $J$ is the recovery current, $a_i, b_i$ are the `electrical' and `mechanical' activation coefficients, $D, \varepsilon$ are coefficients, $U=\Delta \rho$ is the longitudinal density change in lipid layer, $c$ is velocity of unperturbed state in lipid bi-layer, $P, Q$ are the nonlinear coefficients, $H_1, H_2$ are the dispersion coefficients and $\gamma_1, \gamma_2$ are the coupling coefficients for the mechanical wave, $\bar{P}$ is the pressure, $c_{f}$ is the characteristic velocity in the fluid, $\eta_1, \eta_2$ are the coupling coefficients for the pressure wave and $\mu$ is the (viscous) dampening coefficient. `Mechanical' activation coefficients in the action potential and ion current expressions are connected to the improved Heimburg-Jackston part of the model as $b_1 = - \beta_1 U$ and $b_2 = - \beta_2 U$ where $\beta_1, \beta_2$ are the mechanical coupling coefficients. In system~\eqref{EQS} either $J_T$ or $J_X$ are used as coupling forces. 
%
The coupling coefficients are changed for the investigated cases but the rest of the parameters are the same for all the shown solutions. The common parameter values are taken as 
$D=1, \varepsilon=0.01, a_1=0.2, a_2=0.2, c^{2}=0.16,$ $P=-0.05, Q=0.02,\break H_1= 0.43, H_2=0.75, c_{f}^{2}=0.1, \mu=0.0025.$

\subsection{Initial and boundary conditions} 
A $\sech^{2}$-type localized initial condition with an initial amplitudes $Z_o$ and $J_o$ are applied to $Z$ and $J$ in system~\eqref{EQS} and we make use of the periodic boundary conditions for all the members of the model equations
\begin{equation} \label{algtingimus}
\begin{split}
& Z(X,0) = Z_{o} \sech^2 B_{o} X, \quad Z(X,T) = Z (X + 2 K m \pi,T), \quad m = 1,2,\ldots ,\\
& J(X,0) = J_{o} \sech^2 B_{o} X, \quad J(X,T) = J (X + 2 K m \pi,T), \quad m = 1,2,\ldots ,\\
& U(X,0) = 0, \quad U_T(X,0) = 0, \quad U(X,T) = U (X + 2 K m \pi,T), \quad m = 1,2,\ldots ,\\
& \bar{P}(X,0) = 0, \quad \bar{P}_T(X,0) = 0, \quad \bar{P}(X,T) = \bar{P} (X + 2 K m \pi,T), \quad m = 1,2,\ldots ,
\end{split}
\end{equation}
where $K=128$, meaning that the total  length of the spatial period is $256\pi$. The amplitude of the initial condition is taken as $Z_o=2,$ $J_o = 0.1$ and the width parameter is taken as $B_o=1$ for both. In a nutshell -- such an initial condition is a narrow `spark' in the middle of the considered space domain with the amplitude above the threshold resulting in the usual FHN action potential formation which then proceeds to propagate in the positive and negative directions of the 1D space domain under consideration. In the paper only the solutions traveling to the left are shown, i.e., only half the spatial nodes from $0$ to $n/2$. 
%For the amplitude and the width of the initial pulse we use the values $U_{o} = 1$ and $B_{o} = \pi/2$.
%The initial phase speed is taken to be zero, which can be interpreted as starting from the peak of the interaction of two waves propagating in  opposite directions.
For all other equations we take initial excitation to be zero and make use of the same periodic boundary conditions. The solution representing the mechanical and pressure wave is generated over time as a result of coupling with the action potential and ion current parts in the model system. It should be noted that in the present paper wave interactions are not investigated and integration intervals in time are picked such that the waves modeled do not reach the boundaries so the type of boundary conditions used is of low importance. For making use of the pseudospectral method periodic boundary conditions are needed. 

While not shown in the present paper it should be added that the action potentials annihilate each other during the interaction (as expected) but the mechanical and pressure waves can keep on going through many interactions if one uses the fact that we have periodic boundary conditions for taking a look at the interactions of the modeled wave ensembles.

\subsection{The derivatives and integration}
The discrete Fourier transform (DFT) based (PSM) \citep[see][]{Fornberg1998,Salupere2009} is used for numerical solving of the system~\eqref{EQS}. 
Variable $Z$ can be represented in the Fourier space as
\begin{equation} \label{dft}
\widehat{Z}(k,T) = \mathrm{F} \left[ Z \right]= \sum^{n-1}_{j=0}{Z(j \Delta X, T) \exp{\left(-\frac{2 \pi \mathrm{i} j k}{n} \right)}},
\end{equation}
where $n$ is the number of space-grid points ($n=2^{12}$ in the present paper), $\Delta X=2 \pi/n$ is the space step, $k=0,\pm1,\pm2,\ldots,\pm(n/2-1),-n/2$; $\mathrm{i}$ is the imaginary unit, $\mathrm{F}$ denotes the DFT and $\mathrm{F}^{-1}$ denotes the inverse DFT. 
The idea of the PSM is to approximate space derivatives by making use of the DFT 
\begin{equation} \label{dft2}
\frac{\partial^{m} Z}{\partial X^{m}} = \mathrm{F}^{-1}\left[(\mathrm{i} k)^{m} \mathrm{F}(Z) \right],
\end{equation}
reducing therefore the partial differential equation (PDE) to an ordinary differential equation (ODE) and then to use standard ODE solvers for integration with respect to time. The model~\eqref{EQS} contains a mixed derivative term and coupling force terms can be taken either as a space derivative which can be found like in Eq.~\eqref{dft} or time derivative which is not suitable for a direct PSM application and need to be handled separately. 

For integration in time the model system~\eqref{EQS} is rewritten as a system of first order ODE's  after the modification to handle the mixed partial derivative term and a standard numerical integrator is applied. In the present paper ODEPACK FORTRAN code (see \cite{ODE}) ODE solver is used by making use of the F2PY (see \cite{F2PY}) generated Python interface. Handling of the data and initilization of the variables is done in Python by making use of the package SciPy (see \cite{SciPy}).

\subsection{The handling of the mixed derivatives }
Normally the PSM algorithm is intended for $ u_t = \Phi(u,u_x, u_{2x},\ldots,u_{mx})$ type equations. However, we have a mixed partial derivative term $H_2 U_{XXTT}$ in Eqs~\eqref{EQS} and as a result some modifications are needed (see \cite{lauriandrus2009,lauriandruspearu2007,Salupere2009}). %and Publications I--III. %,asktjepp07 
Rewriting system~\eqref{EQS} the equation for $U$ so that all partial derivatives with respect to time are in the left-hand side of the equation 
\begin{equation} \label{LHSofHE}
U_{TT} - H_2 U_{XXTT}= c^{2} U_{XX} + P U U_{XX} + Q U^{2} U_{XX} + P \left( U_{X} \right)^2 + 2 Q U \left(U_X \right)^2 - H_1 U_{XXXX}  + \gamma_1 \bar{P}_T + \gamma_2 J_T
\end{equation}
allows one to introduce a new variable
$\Phi = U - H_2 U_{XX}.$
After that, making use of properties of the DFT, one can express the variable $U$ and its spatial derivatives in terms of the new variable $\Phi$:
\begin{equation}\label{UUXPhi}
U=\mathrm{F}^{-1}\left[\frac{\mathrm{F}(\Phi)}{1+H_2 k^2}\right],
\qquad
\frac{\partial^m U}{\partial X^m}
=\mathrm{F}^{-1}\left[\frac{(\mathrm{i} k)^m \mathrm{F}(\Phi)}{1+H_2 k^2}\right].
\end{equation}
%Here $\mathrm{F}$ denotes the DFT, $\mathrm{F}^{-1}$  the inverse DFT, $k = \pm 1, \pm 2, \ldots \pm(n/2-1), - n/2$, and $n$ is the number of space-grid points.
Finally, in system~\eqref{EQS} the equation for $U$ can be rewritten in terms of the variable $\Phi$ as 
\begin{equation} \label{HEtegelik}
\Phi_{TT} =
c^2 U_{XX} + N U U_{XX} + M U^{2} U_{XX} + N \left( U_{X} \right)^2 + 2 M U \left(U_X \right)^2 - H_1 U_{XXXX}  + \gamma_1 \bar{P}_T + \gamma_2 J_T, 
\end{equation}
where all partial derivatives of $U$ with respect to $X$ are calculated in terms of $\Phi$ by using  expression \eqref{UUXPhi} and therefore one can apply the PSM for numerical integration of Eq.~\eqref{HEtegelik}. Other equations in the  model \eqref{EQS} are already written in the form which can be solved by the standard PSM.

\subsection{The time derivatives $\bar{P}_T$ and $J_T$.}
The time derivatives $\bar{P}_T$ and $J_T$ are found using different methods. For finding $\bar{P}_T$ it is enough to write the equation for $\bar{P}$ in system~\eqref{EQS} as two first order ODE's which is done anyway as the integrator requires first order ODE's and it is possible to extract $\bar{P}_T$ from there directly
\begin{equation} 
\begin{split}
& \bar{P}_{T} = \bar{V} \\
& \bar{V}_{T} = c_{f}^{2} \bar{P}_{XX} + \eta_1 Z_X + \eta_2 J_T - \mu \bar{P}_T.
\end{split}
\label{EQSp}
\end{equation}
For finding $J_T$ a basic backward difference scheme is used
\begin{equation}
J_T (n,T) = \frac{J (n,T) - J (n,(T-dT))}{T - (T - dT)} \approx \frac{\Delta J(n,T)}{d T},
\label{backdiff}
\end{equation}
where $J$ is the ion current from Eqs~\eqref{EQS}, $n$ is the spatial node number, $T$ is the dimensionless time and $dT$ is the integrator internal time step value (which is variable and in the present paper the integrator is allowed to take up to $10^6$ internal time steps between $\Delta T$ values to provide the desired numerical accuracy.

\subsection{The technical details and numerical accuracy}
As noted, the calculations are carried out with the Python package SciPy (see \cite{SciPy}), using the FFTW library (see \cite{FFTW3}) for the DFT and the F2PY (see \cite{F2PY}) generated Python interface to the ODEPACK FORTRAN code (see \cite{ODE}) for the ODE solver. The particular integrator used is the  `vode' with options set to nsteps$=10^6$, rtol$=1e^{-11}$, atol$=1e^{-12}$ and $\Delta T = 2$.

It should be noted that typically the hyperbolic functions like $\sech^2(X)$ in our initial conditions in \eqref{algtingimus} are defined around zero. However, in the present paper the spatial period is taken from $0$ to $K \cdot 2\pi$ which means that the noted functions in \eqref{algtingimus} are actually shifted to the right (in direction of the positive axis of space) by $K \cdot \pi$ so the shape of $\sech^2(X)$ typically defined around zero is actually in our case located in the middle of the spatial period. This is a matter of preference (in the present case the reason is to have more convenient mapping between the values of $X$ and indices) and the numerical results would be the same if one would use a spatial period from $-K \cdot \pi$ to $K \cdot \pi$.

The `discrete frequency function' $k$ in \eqref{dft2} is typically formulated on the interval from $-\pi$ to $\pi$, however, we use a different spatial period than $2\pi$ and also shift our space to be from $0$ to $K \cdot 2\pi$ meaning that 
\begin{equation}
k = \left[\frac{0}{K}, \frac{1}{K}, \frac{2}{K}, \ldots, \frac{n/2 - 1}{K}, \frac{n/2}{K}, - \frac{n/2}{K}, - \frac{n/2 - 1}{K}, \ldots , - \frac{n - 1}{K}, - \frac{n}{K} \right],
\label{diskreetnesagedus}
\end{equation}
where $n$ is number of the spatial grid points uniformly distributed across our spatial period (the size of the Fourier spectrum is $(n/2)$ which is, in essence, the number of spectral harmonics used for approximating the periodic functions and their derivatives) and $K$ is the number of $2\pi$ sections in our space interval. 

There are a few different possibilities for handling the division by zero rising in Eq.~\eqref{backdiff} during the initial initialization of the ODE solver and when the numerical iteration during the integration reaches the desired accuracy resulting in a zero length time step. For initial initialization of the numerical function initial value of $1$ is used for $dT$. This is just a technical nuance as during the initialization the time derivative will be zero anyway as far there is no change in the value of $J(n,0)$. For handing the division by zero during the integration when ODE solver reaches the desired accuracy using values from two steps back from the present time for $J$ and $T$ is computationally the most efficient. Another straightforward alternative is using a logical cycle inside the ODE solver for checking if $dT$ would be zero but this is computationally inefficient. In the present paper a value two steps back in time for calculating $J_T$ is used for all presented results involving $J_T$. The difference between the numerical solutions of the $J_T$ with the scheme using a value 1 step back and additional logic cycle for checking for division by zero and using two steps back in time scheme only if division by zero occurs is only approximately $10^{-6}$ and is not worth the nearly twofold increase in the numerical integration time. 

Overall accuracy of the numerical solutions is approximately $10^{-7}$ for the fourth derivatives, approximately $10^{-9}$ for the second derivatives and approximately $10^{-11}$ for the time integrals. The accuracy of $J_T$ is approximately $10^{-6}$ which is adequate and very roughly in the same order of magnitude as the fourth spatial derivatives. Note that the accuracy estimates are not based on the solving system~\eqref{EQS} with the presented parameters but are based on using the same scheme with the same technical parameters for finding the derivatives of $\sin(x)$ and comparing these to an analytic solution. In addition it should be noted that in the PST the spectral filtering is a common approach for increasing the stability of the scheme -- in the numerical simulations for the present paper the filtering (suppression of the higher harmonics in the Fourier spectrum) is not used although the highest harmonic (which tends to collect the truncation errors from the finite numerical accuracy of floating point numbers in the PST schemes) is monitored as a `sanity check' of the scheme. 
\end{appendices}

%\bibliographystyle{elsarticle-harv}
%\bibliography{library}

\end{document}